\documentclass[preprint,floatfix,nofootinbib]{revtex4-1}
\pdfoutput=1

\usepackage{xcolor}
\usepackage[scaled=1.04]{biolinum} 
\usepackage{fourier}
\usepackage[colorlinks=true,citecolor=teal]{hyperref}
\usepackage{graphicx}
\usepackage{xspace}
\usepackage{amsmath}
\usepackage{amssymb}
\usepackage[babel]{microtype}
\usepackage[english]{babel}
\usepackage[T1]{fontenc}
\usepackage[utf8]{inputenc}
\usepackage{cleveref}
\usepackage{booktabs}
\usepackage{array}
\usepackage{siunitx}
\newcolumntype{C}[1]{>{\centering\let\newline\\\arraybackslash\hspace{0pt}}m{#1}}

\makeatletter
\g@addto@macro\bfseries{\boldmath}
\makeatother

\makeatletter
\let\@afterindenttrue\@afterindentfalse
\@afterindentfalse
\makeatother

\newcommand{\diff}{\,\text{d}}
\newcommand{\given}{\,|\,}
\newcommand{\kl}{D_\text{KL}}
\newcommand{\neff}{n_\text{eff}}
\newcommand{\llangle}{\left\langle}
\newcommand{\rrangle}{\right\rangle}
\newcommand{\pfrac}[2]{\left(\frac{#1}{#2}\right)}
\newcommand{\order}{\mathcal{O}}

\begin{document}

\renewcommand*\abstractname{}

\title{Nested sampling statistical errors}
\author{Andrew Fowlie}
\email{andrew.j.fowlie@njnu.edu.cn}
\author{Qiao Li}
\email{211002028@njnu.edu.cn}
\author{Huifang Lv}
\email{lvhf@njnu.edu.cn}
\author{Yecheng Sun}
\email{07200112@njnu.edu.cn}
\author{Jia Zhang}
\email{jiazhang@njnu.edu.cn}
\author{Le Zheng}
\email{211002054@njnu.edu.cn}

\affiliation{Department of Physics and Institute of Theoretical Physics, Nanjing Normal University, Nanjing, Jiangsu 210023, China}
\begin{abstract}
Nested sampling (NS) is a popular algorithm for Bayesian computation. We investigate statistical errors in NS both analytically and numerically. We show two analytic results. First, we show that the leading terms in Skilling's expression using information theory match the leading terms in Keeton's expression from an analysis of moments. This approximate agreement was previously only known numerically and was somewhat mysterious. Second, we show that the uncertainty in single NS runs approximately equals the standard deviation in repeated NS runs. Whilst intuitive, this was previously taken for granted. We close by investigating our results and their assumptions in several numerical examples, including cases in which NS uncertainties increase without bound.
\end{abstract}

\maketitle

\section{Introduction}

Nested sampling (NS; \cite{skilling}) is a popular Monte Carlo algorithm for Bayesian computation that is widely used throughout physical sciences~\cite{Ashton:2022grj}. NS computes quantities for parameter inference and model comparison simultaneously. For the latter, NS results in an estimate of the evidence integral
\begin{equation}
    Z = \int L(\Theta) \pi(\Theta) \diff \Theta
\end{equation}
where $\Theta$ are a model's parameters, $L(\Theta)$ is the likelihood function, and $\pi(\Theta)$ is the choice of prior. Two models, here labelled $0$ and $1$, may be compared through a so-called Bayes factor \cite{Kass:1995loi}
\begin{equation}
    B_{10} = \frac{Z_1}{Z_0}
\end{equation}
which indicates their relative change in plausibility in light of data.

NS writes evidence integrals using the volume variable $X$
\begin{equation}
    Z = \int_0^1 L(X) \diff X
\end{equation}
where $L(X)$ is the inverse of
\begin{equation}
    X(L^\star) = \int\limits_{L(\Theta) > L^\star} \pi(\Theta) \diff \Theta
\end{equation}
and estimates evidence integrals using statistical estimates of $X(L)$ at known values of $L$. The NS algorithm evolves a collection of $M$ live points. At each iteration, the live point with the worst likelihood, $L^\star$, is replaced by one sampled from the prior subject to the constraint that $L > L^\star$. The volume at iteration $k$ may be estimated from a product of $k$ independent compression factors,
\begin{equation}\label{eq:product_t}
    X_k = \prod_{i=1}^k t_i 
\end{equation}
where the compression factors $t$ are independent and identically distributed, and follow a $\beta(1, M)$ distribution. We may thus estimate $X_k$ through,
\begin{equation}\label{eq:compression}
    \ln X_k \simeq \sum_{i=1}^k \langle \, \ln t_i \rangle = -\frac{k}{M}
\end{equation}

The NS estimates of $Z$ are subject to statistical and systematic errors. The statistical errors originate from the fact that the true $X$ are unknown and estimated statistically and scale as $1/\sqrt{M}$~\cite{Chopin2010,skilling2009nested}. Skilling~\cite{skilling} originally presented an error estimate for $\ln Z$ based on entropy and information theory, though recommended that should be checked using Monte Carlo simulations. Keeton~\cite{keeton}, on the other hand, propagated uncertainties on estimates of $X$ to obtain the variance of estimates of $Z$. 

Keeton~\cite{keeton} found that the two approaches yield remarkably similar numerical answers in many circumstances. This is slightly mysterious, though, as the analytic expressions are quite different and Keeton could only speculate about the cause of the agreement. In \cref{sec:information_theoretic,sec:moments} we simplify the expressions from Skilling and Keeton, respectively, to demonstrate their approximate equivalence analytically. We assume $M \gg 1$ live points and discard terms $\order(1 / M^2)$ and smaller in Keeton's expression for the variance. In \cref{sec:coverage} we explore the coverage properties of the NS uncertainty estimate. We finish in \cref{sec:examples} by applying our results to some numerical problems.

\section{Information theoretic}\label{sec:information_theoretic}

Skilling~\cite{skilling} argues that we can estimate the error in $\ln Z$ by
\begin{equation}\label{eq:skilling_error_ln_z}
    \Delta \ln Z = \sqrt{\frac{\kl}{M}}
\end{equation}
and thus when $\Delta \ln Z \ll 1$,
\begin{equation}\label{eq:skilling_error_ln_z_scaled}
    M \frac{\sigma^2_Z}{Z^2} = \kl
\end{equation}
where $\kl$ is the Kullback-Leibler (KL) divergence~\cite{Kullback1951} between the posterior and prior. This is motivated by considering the number of iterations, $N$, required to reach the posterior bulk, which Skilling assumes lies at about $\ln X = -\kl$,
\begin{equation}
    N = M \kl
\end{equation}
Skilling argues that the uncertainty in $\ln X$ at the moment we reach the posterior bulk dominates the uncertainty in $\ln Z$. From this, we may derive \cref{eq:skilling_error_ln_z} and anticipate that $\ln Z$, rather than $Z$, follows a quasi-Gaussian symmetric distribution.

In general the KL divergence may be written
\begin{equation}
    \kl[p, q] = \int p(x) \ln \pfrac{p(x)}{q(x)}\diff x
\end{equation}
and can be interpreted as a measure of compression. To see this, consider a simple problem: a flat prior, $\pi(\Theta) = \text{const.}$ and a likelihood function, $L(\Theta)$, that vanishes everywhere except a region of volume $\epsilon$, in which it is constant. In this case, $\kl = -\ln \epsilon$. The KL divergence between the posterior and prior may be written using the volume variable $X$ as
\begin{equation}\label{eq:kl_divergence_ns}
    \kl = \int p(X) \ln p(X) \diff X
\end{equation}
where by definition
\begin{equation}
    p(X) \equiv \frac{L(X)}{Z}
\end{equation}
Although the results here make use of the volume variable $X$, they are general and don't assume the use of NS estimators or the NS algorithm.

We introduce the variable $y = -\ln X$ such that
\begin{equation}\label{eq:density_y}
   p(y) = X p(X) = \frac{e^{-y} L(X)}{Z}.
\end{equation}
The KL divergence in \cref{eq:kl_divergence_ns} may be written in the useful form
\begin{equation}
    \kl = - \int p(X) \ln X \diff X + \int p(y) \ln p(y) \diff y
\end{equation}
Thus, the KL divergence is the expected $-\ln X$ minus the (differential) entropy associated with what is known about $y = -\ln X$,
\begin{equation}\label{eq:kl_divergence_ns_expanded}
    \boxed{\kl = \langle -\ln X\rangle  - H(-\ln X)}
\end{equation}
where
\begin{equation}\label{eq:differential_entropy_y}
    H(-\ln X) = -\int p(y) \ln p(y) \diff y.
\end{equation}
This explains the intuition about the posterior bulk lying at $\ln X = -\kl$, as $\kl$ may be written in terms of the expected $-\ln X$. In our analysis, we assume that the moments of $\ln X$ exist; we later check what happens numerically when they do not. We anticipate that usually the second term in \cref{eq:kl_divergence_ns_expanded} would be $\order(1)$,
\begin{equation}
     H(-\ln X) \sim 1.
\end{equation}
In \cref{app:bounds_h_ln_x} we establish bounds on $H(-\ln X) - 1$ such that
\begin{equation}\label{eq:skilling_error_ln_z_scaled_expanded}
    \boxed{M \frac{\sigma^2_Z}{Z^2} \approx - 1 + \langle -\ln X\rangle.}
\end{equation}
The bounds are shown in \cref{eq:skilling_upper_lower}.

\section{Moments of compression factors}\label{sec:moments}

Keeton~\cite{keeton} finds an alternative answer for the error,
\begin{equation}\label{eq:keeton_sigma_2}
    \sigma^2_Z = \langle Z^2 \rangle - \langle Z \rangle^2 = \frac{2}{M(M+1)} \sum_{k=1}^{N} L_k \pfrac{M}{M+1}^k \sum_{i=1}^k L_i \pfrac{M+1}{M+2}^i - \langle Z \rangle^2,
\end{equation}
for $N$ iterations of NS with $M$ live points, where
\begin{equation}
\langle Z \rangle = \sum_{k=1}^N \frac{L_k \pfrac{M}{M+1}^k}{M}.
\end{equation}
Keeton found this by considering the second moments of the compression factors, $\langle X^2\rangle$, and finding the corresponding $\langle Z^2 \rangle$. We may write \cref{eq:keeton_sigma_2} in the form of \cref{eq:skilling_error_ln_z_scaled},
\begin{equation}\label{eq:keeton_error_ln_z_scaled}
    M \frac{\sigma^2_Z}{\langle Z \rangle^2} = \frac{2}{M+1} \sum_{k=1}^{N} \frac{L_k}{Z} \pfrac{M}{M+1}^k \sum_{i=1}^k \frac{L_i}{Z} \pfrac{M+1}{M+2}^i - M
\end{equation}
We will write \cref{eq:keeton_error_ln_z_scaled} as an expectation by defining the discrete probability
\begin{equation}\label{eq:P_k}
P_k = \frac{L_k \pfrac{M}{M+1}^k}{M \langle Z \rangle}
\end{equation}
such that $\sum_{k=1}^N P_k = 1$. This is in fact a discrete form of $p(y)$, i.e., $P_i \simeq p(y) \diff y$ for $y = -\ln X$. We denote the corresponding cumulative mass function by $F_i = \sum_{j=1}^i P_j$. We obtain
\begin{align}
M \frac{\sigma^2_Z}{Z^2} &= 
    \frac{2}{M+1} \sum_{k=1}^{N} {\color{red}\frac{L_k}{Z} \pfrac{M}{M+1}^k} \sum_{i=1}^k {\color{red}\frac{L_i}{Z} \pfrac{M}{M+1}^i} \left(1 + \frac{1}{M^2 + 2M}\right)^i - M \\
    &= \frac{2}{M+1} \sum_{k=1}^{N} {\color{red} M P_k} \sum_{i=1}^k {\color{red} M P_i} \left(1 + \frac{1}{M^2 + 2M}\right)^i - M \\
    &= 2 \sum_{k=1}^{N} P_k \left[ M - 1 + \order\pfrac1M\right] \left[F_k + \sum_{i=1}^k P_i \left(1 + \frac{1}{M^2 + 2M}\right)^i\right] - M \label{eq:keeton_error_as_expectation}
\end{align}
where we also used $M^2 / (M + 1) = M - 1 + \order(1 / M)$. 

We can approximate the second factor in the interior sum using a Taylor expansion,
\begin{equation}\label{eq:expand_keeton_factor}
\left(1 + \frac{1}{M^2 + 2M}\right)^i = 1 + \pfrac{i}{M^2}\left(1 + \order\pfrac{1}{M}\right) + \order\pfrac{i^2}{M^4} \simeq 1  + \frac{i}{M^2}
+ \order\pfrac{i^2}{M^4}
\end{equation}
We neglect terms suppressed by $\order(1 / M)$ and ultimately neglect terms $\order(i^2 / M^4)$ --- that is, we assume $i / M^2 \lesssim 1$. Assuming that $i / M^2 \lesssim 1$ implies that $\Delta \ln Z \lesssim 1$. To see this, first note that $i / M^2$ obtains a maximum at the final iteration, $N / M^2$. Based on Skilling's arguments about compression, or our analysis of the terms in $\kl$, the final iteration should take us beyond $\ln X = - \kl$. Combined with the estimator in \cref{eq:compression}, this implies that $N / M \gg \kl$, and so
\begin{equation}
    \Delta \ln Z \simeq \frac{\kl}{M} \ll \frac{N}{M^2} \lesssim 1
\end{equation}
This is significant as in this regime $\Delta \ln Z \approx \Delta Z / Z$. For well-behaved problems the sum in \cref{eq:keeton_error_ln_z_scaled} may be truncated once most evidence was accumulated, as negligible contributions to the evidence make negligible contributions to the variance of the evidence. Thus the requirement that $M^2 \gtrsim N$ cannot be made arbitrarily severe by running more and more iterations of NS. 

Now let's simplify \cref{eq:keeton_error_as_expectation}. Inside \cref{eq:keeton_error_as_expectation} we have the term $\sum_{k=1}^{N} P_k F_k$. This is the expectation of a cumulative mass function. By connection to the continuous case in \cref{eq:cdf_identity}, we anticipate that this is approximately a half. We may compute it explicitly
\begin{align}\label{eq:expectation_f}
\sum_{k=1}^{N} P_k F_k &= \sum_{k=1}^{N} F_k \Delta F_{k-1} \\
                       &= \frac12 + \frac12 \sum_{k=1}^{N} \Delta F_{k-1}^2 \\
                       &= \frac12 + \frac12 \sum_{k=1}^{N} P_k^2 \\
                       &= \frac12 + \frac12 \frac{1}{\neff}
\end{align}
This is a Riemann sum approximation to the integral of $y = x$. The result differs from one half because the Riemann sum overestimates the integral by sum of triangles lying above the line $y = x$. In the final line, we make use of the effective sample size,
\begin{equation}
\neff = \frac{1}{\sum_{k=1}^{N} P_k^2}
\end{equation}
This gives  
\begin{align}
    M \frac{\sigma^2_Z}{Z^2} &= 2 \sum_{k=1}^{N} P_k  \left[M - 1 + \order\pfrac1M\right] \left[{\color{red}\frac12 + \frac12 \frac{1}{\neff}} + \sum_{i=1}^k P_i \left(\frac{i}{M^2} + \order\pfrac{i^2}{M^4}\right)\right] - M
\end{align}
Finally simplifying,
\begin{equation}
     M \frac{\sigma^2_Z}{Z^2} = - 1 + \frac{M - 1}{\neff} + 2 \sum_{k=1}^{N} P_k\sum_{i=1}^k P_i \left(\frac{i}{M} + \order\pfrac{i^2}{M^3}\right)
    \label{eq:keeton_error_after_cdf_half}
\end{equation}
where we threw away $\order(1 / M)$ terms.

Next we use summation by parts (see \cref{app:sbp}) to re-write the double sum at the end of \cref{eq:keeton_error_after_cdf_half}. We let $P_i = \Delta F_{i-1}$ and denote the interior sum by $I_k$, such that
\begin{align}
        \sum_{k=1}^{N} I_k \Delta F_{k-1} = \left(I_N F_N - I_1 F_0\right) - \sum_{k=1}^{N-1} F_k \Delta I_k
\end{align}
Noting that $I_N = \langle k / M + \order(k^2 / M^3)\rangle$, $F_N = 1$ and that $F_0 = 0$,
\begin{equation}
 = \llangle \frac {k}{M} + \order\pfrac{k^2}{M^3} \rrangle - \sum_{k=1}^{N} F_{k-1} P_{k} \left(\frac{k}{M} + \order\pfrac{k^2}{M^3}\right).
\end{equation}
We write this as
\begin{equation}
    = \frac12 \llangle \frac{k}{M}  + \order\pfrac{1}{M} \frac{k^2}{M^2} \rrangle + \frac12\frac{1}{\neff} - \sum_{k=1}^{N} F_{k-1} P_{k} \left[ \frac{k}{M} - \llangle \frac{k}{M} \rrangle  + \order\pfrac{1}{M} \left(\frac{k^2}{M^2} - \llangle \frac{k^2}{M^2} \rrangle\right)\right]
\end{equation}
where we used 
\begin{equation}
    \sum_{k=1}^{N} F_{k-1} P_k  = \sum_{k=1}^{N} F_{k-1} \Delta F_{k-1} = \frac12 - \frac12 \frac{1}{\neff}
\end{equation}
where we follow similar reasoning as in \cref{eq:expectation_f}. The terms involving $k^2 / M^2$ are related to the mean and variance of $\ln^2 X$. As they are suppressed by $\order(1 / M)$ and we assume $M \gg 1$ we neglect them. Thus \cref{eq:keeton_error_after_cdf_half} becomes
\begin{equation}
M \frac{\sigma^2_Z}{Z^2} = - 1 + \llangle\frac{k}{M}\rrangle + \frac{M}{\neff} - 2 \sum_{k=1}^{N} F_{k-1} P_{k} \left(\frac{k}{M} - \llangle \frac{k}{M} \rrangle\right)
\end{equation}
We may write this result as
\begin{equation}\label{eq:keeton_error_ln_z_scaled_expanded}
     \boxed{M \frac{\sigma^2_Z}{Z^2}= - 1 + \llangle-\ln X\rrangle + \frac{M}{\neff} - 2 \sum_{k=1}^{N} F_{k-1} P_{k} \left(\frac{k}{M} - \llangle \frac{k}{M} \rrangle\right)}
\end{equation}
where we used $\llangle k / M\rrangle$ as an estimator of the expectation $\langle -\ln X\rangle$ since $k / M$ is an unbiased estimator of $-\ln X_k$. The first two terms, $-1 + \langle -\ln X\rangle$, exactly match those in \cref{eq:skilling_error_ln_z_scaled_expanded}. We bound the final two terms in \cref{app:bound_neff,app:bound_sum}, establishing the approximate equivalence of Skilling's and Keeton's error formulae in \cref{eq:skilling_error_ln_z_scaled,eq:keeton_error_ln_z_scaled}. The term $M / \neff$ can be no more than than about a half. The magnitude of the final term is bounded by the standard deviation of $- \ln X$, $\sigma$. Skilling justified \cref{eq:skilling_error_ln_z} by arguing that $-\ln X$ was typically peaked about some region, in which case we expect $\sigma \ll \langle -\ln X \rangle$. 

\section{Frequentist coverage}\label{sec:coverage}

The error estimates that we considered represent the uncertainty in a single run. The likelihoods, $L$, are known, but the volumes, $X$, are uncertain. That is, the second factor in red isn't known, 
\begin{equation}\label{eq:z_single_run}
    Z = \sum_{i=1}^N L_i \, \Delta {\color{red}X_i} = \sum_{i=1}^N L(X(L_i)) \, \Delta {\color{red}X_i}
\end{equation}
We estimate $X_i$ from the statistics of the NS procedure and \cref{eq:product_t}. We denote this distribution by $X_i \sim \prod_i\beta(1, M)$.

In repeated NS runs with fixed estimators for the volumes, $\hat X$, it is the likelihoods associated with each volume that change. That is, it is the first factor in red changes between runs
\begin{equation}
    Z = \sum_{i=1}^N {\color{red}L_i} \, \Delta \hat X_i = \sum_{i=1}^N {\color{red}L(X_i)} \, \Delta \hat X_i
\end{equation}
where as before $X_i \sim \prod_i\beta(1, M)$. Making use of first order Taylor expansions of $L(X)$,
\begin{align}
    Z &\approx \sum_{i=1}^N L(\hat X_i) \Delta \hat X_i + L^\prime(\hat X_i) (X_i - \hat X_i) \Delta \hat X_i\\
    &=\sum_{i=1}^N L(\hat X_i) \Delta \hat X_i - (L(\hat X_i) - L(\hat X_{i+1})) (X_i - \hat X_i)\\
    &=\sum_{i=1}^N L(\hat X_i) \Delta \hat X_i - \Delta L(\hat X_i) (X_i - \hat X_i)\\
    &=\sum_{i=1}^N L(\hat X_i) \Delta \hat X_i + \sum_{i=1}^{N-1} L(\hat X_{i+1}) (\Delta X_i - \Delta \hat X_i)
\end{align}
where we used summation by parts in \cref{app:sbp} and the boundary terms are zero. Up to quadrature errors, we see that there is a cancellation such that
\begin{align}
    Z = \sum_{i=1}^N L(\hat X_i) \Delta X_i
\end{align}
and thus it varies in the same way as \cref{eq:z_single_run}. Since this required a first-order Taylor expansion of $L(X)$, we require $L(X)$ to be approximately linear on scales $\Delta X$ and $\hat X - X$, which are typically about $X / M$.  

\section{Examples and numerical checks}\label{sec:examples}

We turn our consideration to numerical examples. Our analysis used the mean and variance of $\ln X$; we thus consider likelihood functions that result in heavy-tailed and multi-modal distributions in $p(\log X)$, including ones in which the mean or variance of $\ln X$ don't exist. The problems are detailed in \cref{app:problems}.

For each problem, we compute 
\begin{itemize}
    \item The distribution of $\ln Z$ in repeated calculations
    \item The distribution of $\ln Z$ in single calculations from simulating the compression factors
    \item The error estimates and their constitute parts, such as $H(\ln X)$, $\kl$, $\langle - \ln X \rangle$, and the variance of $\ln X$
\end{itemize}
We show our results in \cref{tab:results}. Our final two toy problems pose problems for conventional automatic stopping criteria in NS. The likelihood is unbounded from above and so it is easy to overestimate the remaining evidence and continue integration. We sidestep this issue by halting integration manually after an $N = 50 M$ iterations. 

\begin{figure}[th]
    \centering
    \includegraphics[width=1\textwidth]{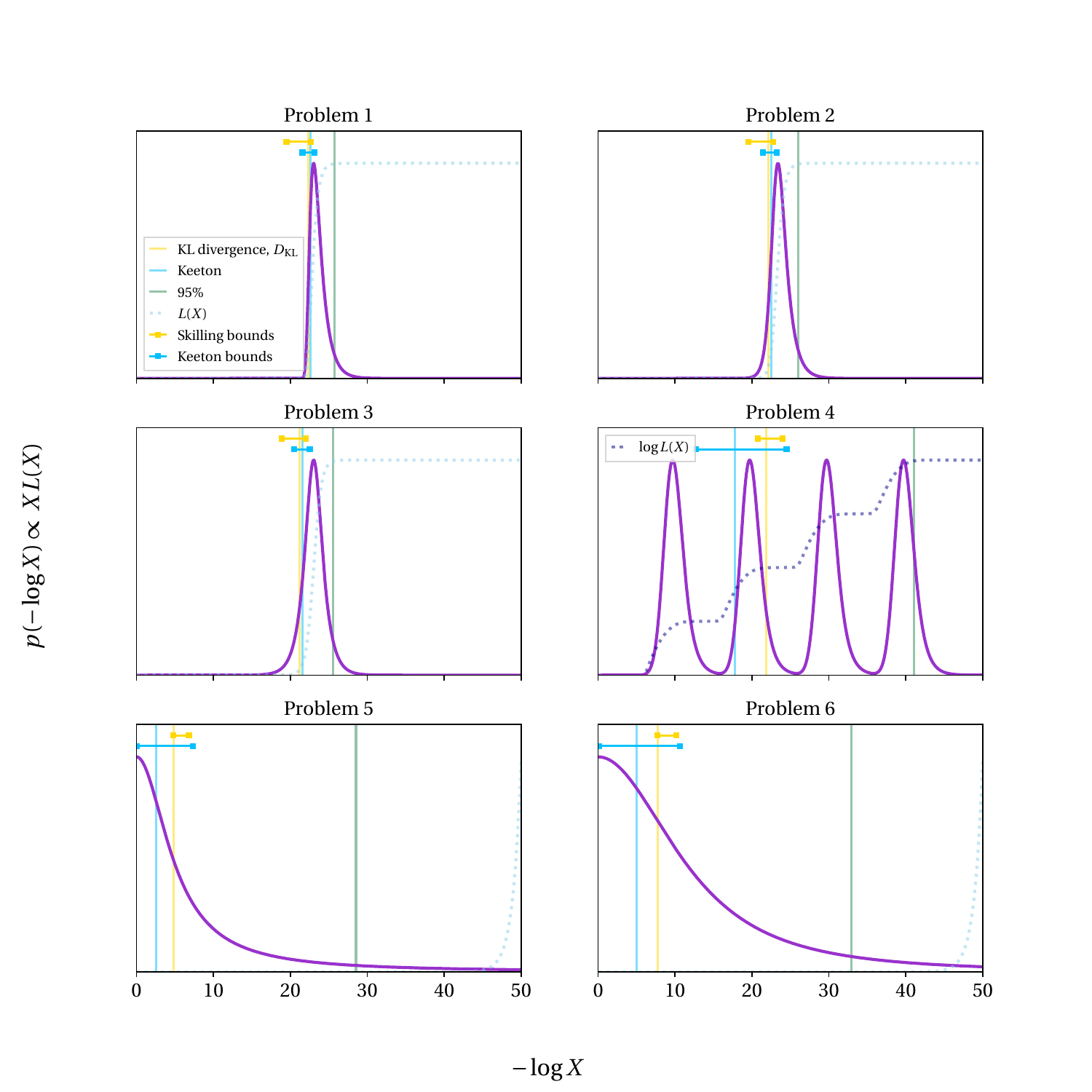}
    \caption{Distributions $p(-\ln X)$ for our six toy problems (solid lines) and likelihoods $L(X)$ (dashed lines). We show Skilling and Keeton error estimates (vertical lines) as well as the region containing $95\%$ of the posterior mass. The error bars show the bounds in \cref{eq:skilling_upper_lower,eq:keeton_error_bound}.}
    \label{fig:dist_ln_x}
\end{figure}

The distributions $p(y = -\ln X)$, critical to our analysis, are shown in \cref{fig:dist_ln_x}. We illustrate Skilling's and Keeton's error estimates by plotting $M \sigma_Z^2 /Z^2$
as vertical lines. We take $L_i = L(\hat X_i)$ in this plot and stop at $N = 50 M$ iterations. We furthermore show the bounds that we established in \cref{app:bounds_h_ln_x,app:bound_neff,app:bound_sum}. In the first three toy problems, the Keeton and Skilling estimates lie in close agreement. In the fourth problem in which $p(\log X)$ was multimodal, they disagree somewhat. They may disagree in this case as the variance of $\ln X$ is moderate, owing to the multiple modes the posterior at different depths in $\ln X$. This moderate variance expands the allowed interval of the Keeton estimator towards zero. In such cases, the Skilling estimate is likely to be conservative and over-estimate the error, as in this case. This repeats in the fifth and sixth problems, heavy-tailed cases in which the variance of $\ln X$ in fact diverges.

\begin{table}[th]
    \centering
    \begin{tabular}{C{2cm}SSSS}
    \toprule
    & \multicolumn{2}{c}{Simulations} & \multicolumn{2}{c}{Analytic} \\
    \midrule
    {Problem} & {Standard deviation} & {Uncertainty} & {Skilling} & {Keeton} \\
    \midrule
1 & 0.14753376667974746 & 0.15396970799078974 & 0.14929189463700254 & 0.15034072211265564 \\
2 & 0.14765662459870452 & 0.15245625199555934 & 0.14864106960897583 & 0.14993471456082733 \\
3 & 0.14491878266695388 & 0.14646422098154202 & 0.14553112967131168 & 0.14685028708898226 \\
4 & 0.13275035901101556 & 0.1306369330211966 & 0.1481077330641047 & 0.13359017011333688 \\
5 & 0.07044842603710902 & 0.07444222899412883 & 0.08839878392801767 & 0.07070518735449292 \\
6 & 0.05068328898663434 & 0.061541437971821045 & 0.06977210269589536 & 0.05113753183978759 \\
    \bottomrule
    \end{tabular}
    \caption{The uncertainty in $\ln Z$ in our six toy problems. We find the uncertainty through analytic expressions and simulations of $\ln Z$. For the latter, first we simulate $\ln Z$ through repeated NS runs using fixed estimators of the volume, $\hat X$ and find the standard deviation. Second we simulate $\ln Z$ through simulations of the volume variable using fixed likelihood levels found from a single NS run.}
    \label{tab:results}
\end{table}

Our numerical results are shown in \cref{tab:results}. As well as the uncertainty estimates, we show the standard deviation of results from repeated NS runs using fixed estimators of the volume, $\hat X$, and from repeated simulations of the volumes for a single NS run. In each case we used $M = 1000$ live points and found standard deviations from $10\,000$ repeats. We see close similarly between estimators and simulations in all cases.

Remarkably, Skilling's estimator performs reasonably well even in the sixth problem for which $\kl$ diverges. The divergence occurs in the tail of the integral in $y = -\log X$. In NS, we integrate from $y = 0$ to $y =\infty$ in steps of about $1 / M$. We stop after $N$ iterations reaching about $y \approx N / M$. We assume that at this point the evidence was accumulated with negligible truncation error. As we must stop integrating, we never see the divergences, and Skilling and Keeton are applied to a truncated problem with finite moments of $y$ and finite KL divergence. As we increase $N$, though, we see more of the tail and the moments grow. As the mean and variance of $y$ grow, the difference between Skilling and Keeton's estimates allowed by our formulas grows.

In fact, further consideration suggests that Keeton's error estimate in \cref{eq:keeton_error_ln_z_scaled} diverges for in the fifth and sixth problems in the limit $N \to \infty$ for fixed $M$. We may bound the error by including only the final term in the interior sum,
\begin{align}
    M \frac{\sigma^2_Z}{\langle Z \rangle^2} &\le \frac{2}{M+1} \sum_{k=1}^{N} \frac{L_k}{Z} \pfrac{M}{M+1}^k \frac{L_k}{Z} \pfrac{M+1}{M+2}^k\\
    &= \frac{2}{M+1} \sum_{k=1}^{N} \frac{L^2_k \hat X_k^2}{Z^2} \left(1 + \frac{1}{M^2 + 2M}\right)^k
\end{align}
This may diverge. For example, for the sixth problem and using $L_k = L(\hat X_k)$ the bound would be proportional to the sum
\begin{align}
    \sum_{k=1}^{N} \frac{1}{(\beta^2 + k^2)^2} \left(1 + \frac{1}{M^2 + 2M}\right)^k
\end{align}
for $\beta = \text{const.}$ This diverges. Thus paradoxically, running NS for longer increases the error without bound, despite the fact that  the integral is finite and that most mass was already accumulated. 

\begin{figure}[th]
    \centering
    \includegraphics[width=1\textwidth]{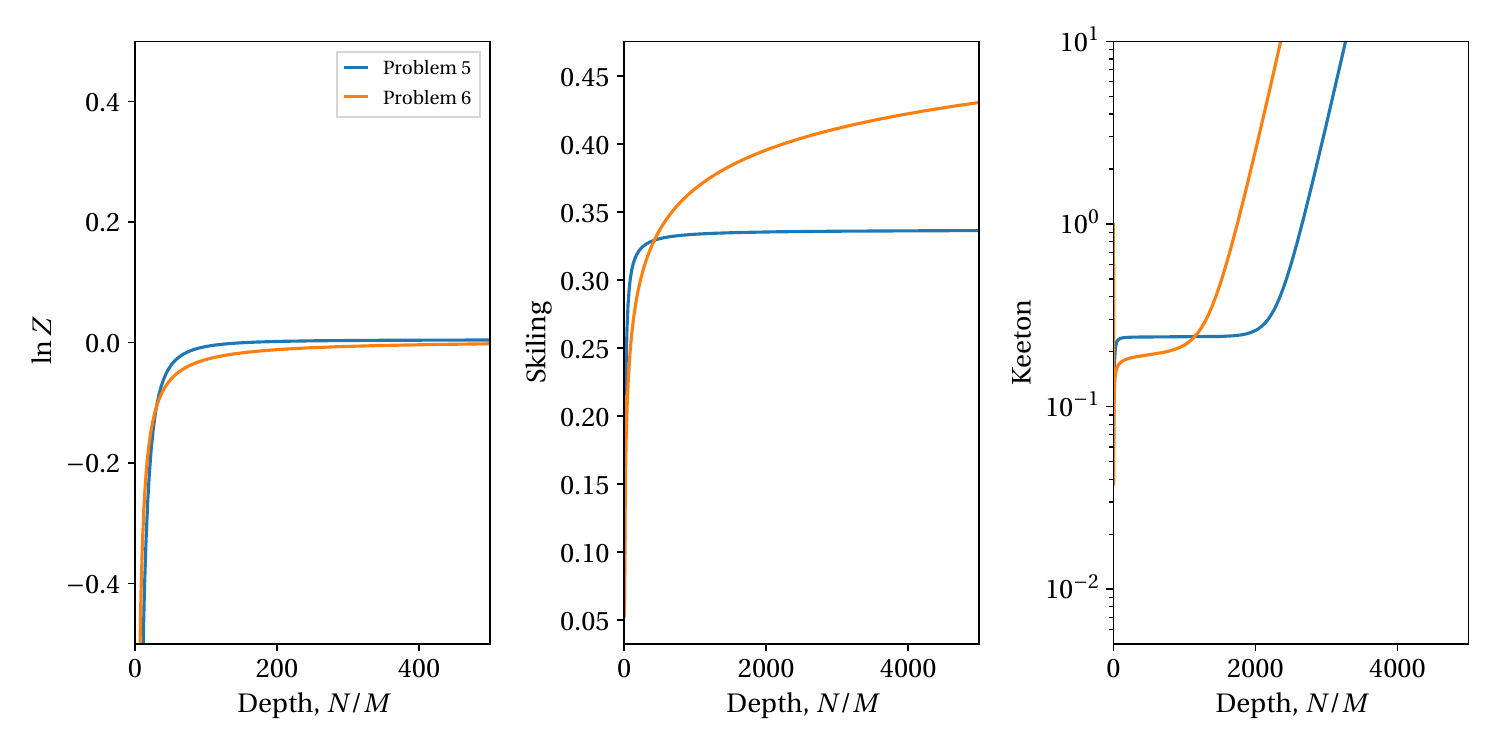}
    \caption{NS results with $L_i = L(\hat X_i)$ for problems five and six as we increase the number of iterations, $N$, for $M=100$ live points. We show $\ln Z$ as well as Skilling's and Keeton's error estimates.}
    \label{fig:n_iter}
\end{figure}

We explore this phenomena in \cref{fig:n_iter} by showing NS results as we increase $N$ in problems five and six when $L_i = L(\hat X_i)$ and for $M = 100$ live points. The growth in the error in problems five and six may be partly understood by the fact that these problems are especially pathological. For example, for $X \sim U(0, 1)$ the variance of $L(X)$ in problems five and six diverges such that the variance of a Monte Carlo estimate of the evidence integral would diverge. The divergence in the variance originates from the $X = 0$ singularity in $L(X)$ and we approach it in NS as we increase the number of iterations. In problem five Skilling's error asymptotes whereas Keeton's slowly diverges whereas in problem six both estimates quickly diverge.

\section{Summary}

We demonstrated that the dominant terms in Skilling's and Keeton's expressions for the statistical uncertainty in NS estimates of the evidence are both $-1 + \langle -\ln X \rangle$, assuming that $\Delta Z / Z \ll 1$. This explains the numerical agreement between them which was previously somewhat mysterious. We cross-checked our analytic findings across six toy problems, including pathological cases with phase transitions and cases in which the moments of $\ln X$ didn't exist. We showed that in well-behaved problems Skilling's and Keeton's estimates are reliable and in agreement with simulations. In pathological cases, however, may drive an arbitrary distance between them by, for example, increasing the variance of the posterior distribution of $-\ln X$, though Keeton's estimate remained in agreement with simulations.  This validates the intuition that Skilling's estimate would apply in cases in which $-\ln X$ contained a single narrow peak. Lastly, we explored cases in which the uncertainty diverges as the number of iterations increases. This is a previously unknown weakness of the NS algorithm. 

\bibliographystyle{JHEP}
\bibliography{sample}

\renewcommand{\appendixname}{APPENDIX}
\appendix

\section{Properties of differential entropy}

\subsection{Definitions and notation}

Differential entropy
\begin{equation}
    H(x) \equiv - \int p(x) \ln p(x) \diff x
\end{equation}
The notation may be misleading as this isn't a function of $x$; it's a functional of $p$. Thus where convenient we sometimes denote this as,
\begin{equation}
    H[p] \equiv - \int p(x) \ln p(x) \diff x
\end{equation}
in this case we use square brackets. 

The conditional differential entropy,
\begin{equation}
    H(x \given y) \equiv \int H\left[p(x \given y)\right] p(y) \diff y
\end{equation}
where
\begin{equation}
    p(x \given y) = \frac{p(x, y)}{p(y)}
\end{equation}

The KL divergence
\begin{equation}
    \kl[p, q] \equiv \int p(x) \ln \pfrac{p(x)}{q(x)} \diff x
\end{equation}

\subsection{Concavity of the differential entropy}\label{sec:concave_differential_entropy}

First, note that
\begin{equation}
    H(x \given y) = -\kl\left[p(x, y), p(x) p(y)\right] + H(x)
\end{equation}
It is well-known that $\kl \ge 0$ --- this is Gibbs' inequality --- such that
\begin{equation}
    H(x) \ge H(x \given y)
\end{equation}

Now we write without loss of generality
\begin{equation}
    p(x) = \int p(x \given y) p(y) \diff y
\end{equation}
and we have
\begin{equation}
    H[p(x)] \ge \int H\left[p(x \given y)\right] p(y) \diff y
\end{equation}
which is all we need.

\section{Summation by parts}\label{app:sbp}

This is an analogue of integration by parts,
\begin{equation}
    \int f \diff g = fg - \int g \diff f
\end{equation}
For sums, we have that
\begin{equation}
    \sum_{i=m}^n f_i \Delta g_i = \left(f_n g_{n+1} - f_m g_m\right) - \sum_{i=m}^{n-1} g_{i+1} \Delta f_i
\end{equation}
where $\Delta f_i = f_{i+1} - f_i$.

\section{Expectation of maximum}\label{sec:expectation_maximum}
Consider
\begin{align}
    \langle \max(0, y) \rangle \equiv \int p(y) \max(0, y) \diff y
\end{align}
for $\langle y \rangle = 0$.
First note that
\begin{equation}
    \max(0, y) = \frac{y + |y|}{2}
\end{equation}
such that 
\begin{align}
    \langle \max(0, y) \rangle = \frac12\langle |y| \rangle
\end{align}
Now note that
\begin{equation}
    \langle |y| \rangle \le \sqrt{\llangle |y|^2 \rrangle} = \sigma
\end{equation}
by Jensen's inequality since squaring is convex. Putting things together,
\begin{equation}
    \boxed{\langle \max(0, y) \rangle \le \frac12 \sigma}
\end{equation}
when $\langle y \rangle  = 0$.

\section{Expectation of cumulative density function}

For any distribution $p$ with CDF $F$, such that $p(x) = \text{d}F / \text{d}x$,
\begin{equation}\label{eq:cdf_identity}
    \langle F \rangle = \int p(x) F(x) \diff x = \int_0^1 F \diff F = \frac12
\end{equation}
This is intuitive --- the expectation of a CDF is a half.

\section{Bounds on $H(-\ln X)$}\label{app:bounds_h_ln_x}

For fixed expectation, $\langle x \rangle = \text{const.}$, the exponential distribution
\begin{equation}
    p(x) = \lambda e^{- \lambda x}
\end{equation}
where $1/\lambda = \langle x \rangle$ maximises the entropy for a positive random variable~\cite{https://doi.org/10.1111/j.1467-9574.1972.tb00152.x}. The differential entropy of this distribution is $1 - \ln \lambda = 1 + \ln \langle x \rangle$. Applying this result to the positive random variable $y = -\ln X$ with fixed expectation $\langle y \rangle = \langle - \ln X \rangle$ results in an upper bound on the differential entropy,
\begin{equation}\label{eq:upper_bound_differential_entropy_y}
H(-\ln X) \le 1 + \ln \langle -\ln X\rangle
\end{equation}
since $1 + \ln \langle -\ln X\rangle$ must be the maximum.

For a lower bound, consider the fact that any monotonic function may be written as a sum or integral of step functions. As $L(X)$ is a monotonically decreasing function in NS,\footnote{Although $L(X)$ is increasing over an NS run, as $X$ goes from $1$ to $0$ and so $L(X)$ is a monotonically decreasing function.}
we can write it as
\begin{equation}\label{eq:monotonic_representation_likelihood}
L(X) = \int_0^1 w(u) \, \Theta(u - X) \diff u
\end{equation}
where $\Theta$ is a step function and $w(u) \ge 0$. This would lead to a distribution for $y = -\ln X$ from \cref{eq:density_y}
\begin{equation}
p(y) = \int w(u) \frac{e^{-y} \, \Theta(u - e^{-y})}{Z} \diff u
\end{equation}
We can in fact write this as an integral over a marginal distribution
\begin{equation}\label{eq:density_y_mixture}
    p(y) = \int p(y \given u) p(u) \diff u
\end{equation}
where
\begin{align}
    \label{eq:density_y_conditional}
    p(y \given u) &= \frac{e^{-y} \, \Theta(u - e^{-y})}{u}\\
    p(u) &= \frac{u w(u)}{Z}
\end{align}
with $\int p(u) \diff u = 1$ and $\int p(y \given u) \diff y= 1$.

Because entropy is concave (see \cref{sec:concave_differential_entropy}), combining distributions leads to an entropy that is bigger than the sum of the individual entropies. That means for our mixture in \cref{eq:density_y_mixture}
\begin{equation}
H(-\ln X) = H(y) \ge \int p(u) H\left[p(y \given u)\right] \diff u
\end{equation}
We can evaluate the differential entropy $ H\left[p(y \given u)\right]$ appearing inside that integral --- it's independent of $u$. \Cref{eq:density_y_conditional} is just an exponential distribution with $\lambda = 1$ shifted by $\ln u$. Shifting a distribution doesn't change its entropy, which for $\lambda = 1$ is $H = 1$,
\begin{equation}
    H\left[p(y \given u)\right] = 1 \quad\text{for all $u$}.
\end{equation}
Thus we get 
\begin{equation}\label{eq:lower_bound_differential_entropy_y}
H(-\ln X) \ge \int p(u) 1 \diff u = 1.
\end{equation}
Combining \cref{eq:lower_bound_differential_entropy_y,eq:upper_bound_differential_entropy_y} gives
\begin{equation}\label{eq:skilling_upper_lower}
\boxed{1  \le H(-\ln X) \le 1 + \ln \langle -\ln X\rangle}
\end{equation}

For these bounds to never be in conflict, we require $\ln \langle -\ln X\rangle \ge 0$ and so $\langle -\ln X\rangle \ge 1$. We can prove that this is indeed a bound by again making use of the representation of $L(X)$ in \cref{eq:density_y_mixture},
\begin{align}
    \langle -\ln X\rangle &= \int_0^1 p(u) \left[\int_0^\infty p(y \given u) y \diff y \right]\diff u\\
    &= \int_0^1 p(u) \left[\int_{-\ln u}^\infty \frac{y e^{-y}}{u} \diff y\right] \diff u \\
   &= \int_0^1 p(u) \left[1 - \ln u\right] \diff u
\end{align}
The factor in square brackets is a monotonically decreasing function of $u$. Thus the minimum occurs at $u=1$ and so the minimum $\langle -\ln X\rangle = 1$ occurs when $p(u) = \delta(u - 1)$. This corresponds to $L(X) = \text{const.}$

\section{Bounds on $\neff$}\label{app:bound_neff}

We wish to bound the effective sample size, $\neff$. In general, we could minimize it by assinging all weight to a single sample. In the context of NS, the closest we can get to that is to assign samples zero weight everywhere except a region $X \le f$, that is, $L(X) \propto \Theta(f - X)$. Explicit computation gives
\begin{equation}
\neff \ge 2 M + 1
\end{equation}
For an upper bound on the effective sample size, consider that by Jensen's inequality,
\begin{equation}
\frac{1}{\neff} = \sum_{k=1}^{N} P_k P_k = \langle P \rangle \ge e^{\llangle \ln P \rrangle}
\end{equation}
In fact, the right-hand side is the reciprocal of the channel capacity, which is another estimator of sample size. We find that,
\begin{align}
\frac{1}{\neff} \ge e^{\llangle \ln P \rrangle} &= \frac{1}{M} e^{\sum_{k=1}^{N} P_k \ln(P_k M)}\\
                                                &= \frac{1}{M} e^{\sum_{k=1}^{N} P_k \ln(P_k \Delta \ln X_k)}\\
                                                &= \frac{1}{M} e^{-H(-\ln X)}
\end{align}
In the final line, we used the fact that the sum estimates the differential entropy in \cref{eq:differential_entropy_y}. Thus
\begin{equation}
\neff \le M e^{H(-\ln X)} \le M e \, \langle -\ln X \rangle
\end{equation}
The final inequality comes from the bound established in \cref{eq:upper_bound_differential_entropy_y}.

\section{Bounds on the remaining sum}\label{app:bound_sum}

We may write the final term in \cref{eq:keeton_error_ln_z_scaled_expanded} as
\begin{equation}
    2 \sum_{k=1}^{N} P_k F_{k-1} z_k
\end{equation}
where $z_k = k/M - \langle k / M\rangle$. We may now construct the bounds,
\begin{align}
    0 &\le 
    2 \sum_{k=1}^{N} P_k F_{k-1} z_k\\
    &\le 
    2 \sum_{k=1}^{N} P_k F_{k-1} \max(0, z_k)\\ 
    &\le 
    2 \sum_{k=1}^{N} P_k \max(0, z_k)  \\
    &=
    2 \langle \max(z, 0) \rangle \le \sigma 
\end{align}
where $\sigma$ is the standard deviation of $z$ as well as of $y$. The lower bound of $0$ comes from the fact that $\langle z \rangle = 0$, and by multiplying by the monotonically increasing and positive $F_{k-1}$ surely we're favouring larger $z$ than previously. The second inequality comes from the fact $\max(0, z_k) \ge z_k$. The third inequality comes from the fact that $F_{k-1} \le 1$. For the final inequality, we used the inequality in \cref{sec:expectation_maximum}. 
Thus we find
\begin{equation}\label{eq:keeton_error_bound}
 0 \le 2 \sum_{k=1}^{N} P_k F_{k-1} z_k \le \sigma
\end{equation}
where $\sigma$ is the standard deviation of $y = -\ln X$. 

\section{Toy problems}\label{app:problems}

We construct one-dimensional toy problems. For each problem, the prior is simply a uniform $\mathcal{U}(0, 1)$,
\begin{equation}
    \pi(x) = \begin{cases} 
    1 & 0 \le x \le 1 \\
    0 & \text{elsewhere}
    \end{cases}
\end{equation}
The likelihood function, $L(x)$, is a monotonically decreasing function such that $L(x)$ and the overloaded $L(X)$ are truly the same function.

We consider six problems:
\begin{enumerate}
\item A one-sided Gaussian density,
\begin{equation}
L(x) = \frac{2}{\sqrt{2\pi}\sigma} e^{- \frac{x^2}{2\sigma^2}}
\end{equation}
normalised such that $Z = 1$.
We choose $\sigma = 10^{-10}$. Analytically,
\begin{equation}
    \kl = -\ln\sigma - \frac12 - \ln \frac{2}{\sqrt{2\pi}}
\end{equation}
and
\begin{equation}
    \langle -\ln X \rangle = -\ln \sigma + \frac12 (\gamma_E + \ln 2) 
\end{equation}
where $\gamma_E = 0.577\ldots$ is the Euler-Mascheroni constant.

\item A one-sided student's $t$ with two degrees of freedom and a scale parameter
\begin{equation}
    L(x) = \frac{\gamma^2}{(\gamma^2 + x^2)^{3/2}}
\end{equation}
normalised such that $Z = 1$.
We choose $\gamma = 10^{-10}$. Analytically,
\begin{equation}
   \kl = -\ln \gamma + 3(\ln 2 - 1)
\end{equation}
and 
\begin{equation}
   \langle -\ln X \rangle = -\ln\gamma + \ln 2
\end{equation}

\item A one-sided Cauchy density
\begin{equation}
    L(x) = \frac{2}{\pi} \frac{\gamma}{\gamma^2 + x^2}
\end{equation}
normalised such that $Z = 1$.
We choose $\gamma = 10^{-10}$. Analytically,
\begin{equation}
    \kl = - \ln\gamma - \ln2\pi
\end{equation}
and
\begin{equation}
    \langle -\ln X \rangle = - \ln\gamma - \sqrt{\frac{2}{\pi}} G
\end{equation}
where $G= 0.916\ldots$ is Catalan's constant.

\item A likelihood that exhibits phase transitions,
\begin{equation}\label{eq:likelihood_pt_problem}
L(x) = \sum_i e^{\mu_i} \, \Phi(-\ln x; \mu_i, \sigma^2 = 1)
\end{equation}
where $\Phi$ is a normal cumulative density function. We choose $\mu = \{10, 20, 30, 40\}$. Each term in the sum produces a plateau in the likelihood function that corresponds to a phase. 

We show the phase transition phenomena in \cref{fig:pt}. In annealing methods the likelihood is raised to the inverse temperature, $L(X)^\beta$, and we cool from $\beta = 0$ to $\beta = 1$. The modes in the distribution of $-\ln X$ change rapidly with temperature as we cool and evolving samples such that all modes are populated could be challenging.

\begin{figure}
    \centering
    \includegraphics[width=0.7\textwidth]{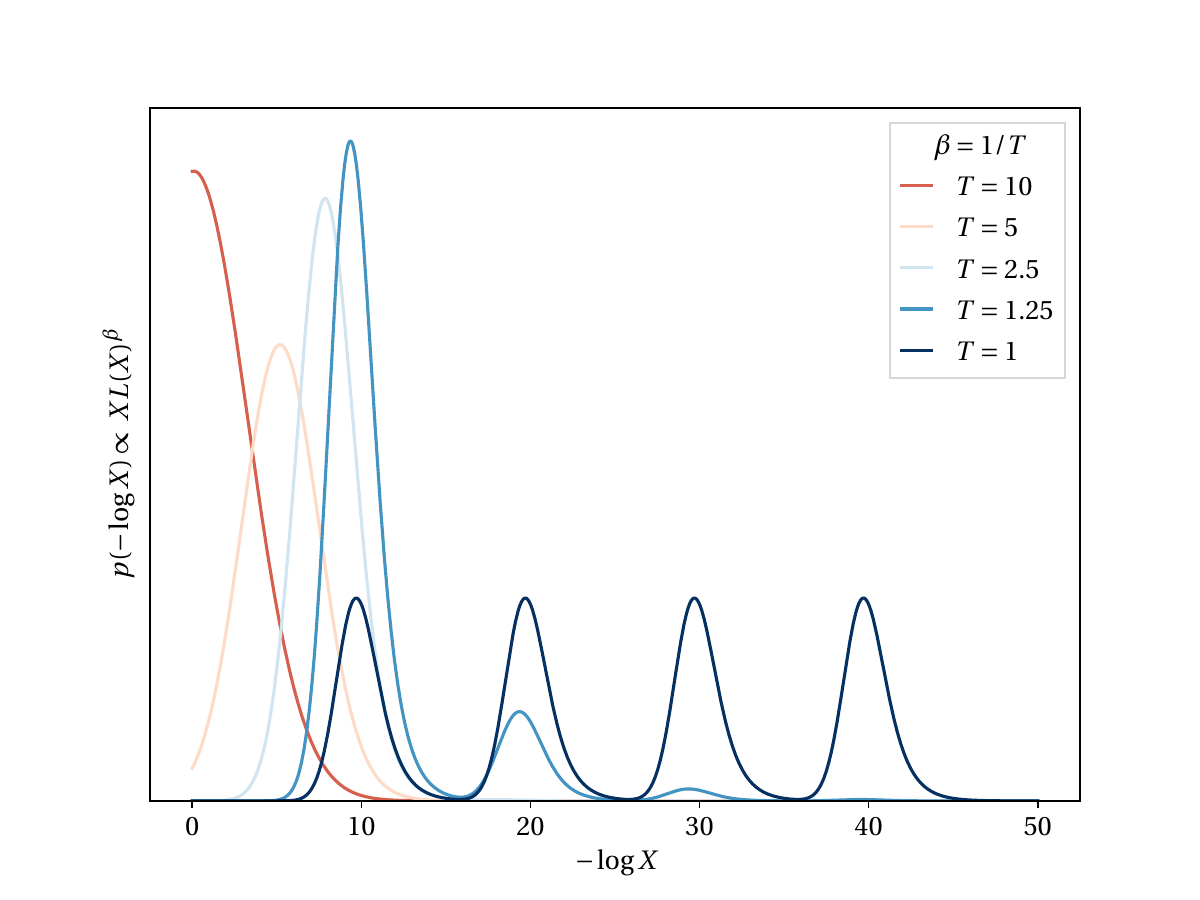}
    \caption{Our phase transition problem in \cref{eq:likelihood_pt_problem}.}
    \label{fig:pt}
\end{figure}

\item A one-sided Log-student's $t$ density with two degrees of freedom and a scale parameter,
\begin{equation}
    L(x) = \frac{1}{x} \frac{\gamma^2}{(\gamma^2 + \ln^2 x)^{3/2}}
\end{equation}
with $\gamma \ge 1$ and normalised such that $Z = 1$. We choose $\gamma = 15$.
The likelihood blows up at $x = 0$ like $1 / (x \ln^3 x)$ though the evidence integral remains finite as $p(-\ln X)$ follows a student's $t$ on $(0, \infty)$. The variance of $\ln X$ diverges. Analytically,
\begin{equation}
    \kl = \gamma - \ln \gamma + 3(\ln 2 - 1)
\end{equation}
and
\begin{equation}
    \langle -\ln X \rangle = \gamma
\end{equation}

\item A one-sided Log-Cauchy density,
\begin{equation}
    L(x) = \frac{1}{x} \frac{2}{\pi} \frac{\gamma}{\gamma^2 + \ln^2 x_i}
\end{equation}
with $\gamma \ge 1$ and normalised such that $Z = 1$. We choose $\gamma = 5$.  
The likelihood blows up at $x = 0$ like $1 / (x \ln^2 x)$ though the evidence integral remains finite as $p(-\ln X)$ follows a Cauchy on $(0, \infty)$. The moments of $\ln X$ and the KL divergence diverge, though $H(\ln X) = \ln(2\pi\gamma)$. 

\end{enumerate}

In the final two problems, $p(y = - \ln X)$ is monotonically decreasing --- it doesn't have a peak. However, problems with a peak and but with otherwise similar properties can easily be engineered. For example, we could multiply by the likelihood functions in the final two problems by a normal cumulative density function.

Lastly, note that we cannot construct pathological cases in which $\langle - \ln X\rangle$ diverges whilst the KL divergence remains finite; if $\langle - \ln X\rangle$ diverges, so does $\kl$. To see this, we can write
\begin{equation}\label{eq:kl_y}
    \kl = \int_0^\infty p(y) (y + \ln p(y)) \diff y.
\end{equation}
where $y = -\ln X$. For 
\begin{equation}
    \langle - \ln X\rangle = \int_0^\infty p(y) y \diff y
\end{equation}
to diverge, $p(y)$ must go to zero no faster than $1/y^2$. Such that in the factor $y + \ln p(y)$ in \cref{eq:kl_y}, the term $\ln p(y)$ cannot stop the divergence, as it grows no faster than logarithmically.

\end{document}